\documentclass[lettersize,journal]{IEEEtran}
\usepackage{amsmath,amsfonts}
\usepackage{algorithmic}
\usepackage{algorithm}
\usepackage{array}
\usepackage{textcomp}
\usepackage{stfloats}
\usepackage{url}
\usepackage{CJKutf8}
\usepackage{verbatim}
\usepackage{graphicx}
\usepackage{cleveref}
\usepackage{cite}
\usepackage{utfsym}
\usepackage{subcaption}
\usepackage{booktabs}
\usepackage{mhchem}

\begin{document}
\begin{CJK}{UTF8}{gbsn}
\title{Filtering Reconfigurable Intelligent Computational Surface for RF Spectrum Purification}

\author{Kaining Wang, Bo~Yang, Zhiwen Yu, Xuelin~Cao,   
 M\'erouane Debbah,
 and Chau Yuen
\thanks{K. Wang, B. Yang, and Z. Yu are with the School of Computer Science, Northwestern Polytechnical University, Xi'an, Shaanxi, 710129, China. 

X. Cao is with the School of Cyber Engineering, Xidian University, Xi'an, Shaanxi, 710071, China. 

M. Debbah is with the Center for 6G Technology, Khalifa University of Science and Technology, P O Box 127788, Abu Dhabi, United Arab Emirates. 

C. Yuen is with the School of Electrical and Electronics Engineering, Nanyang Technological University, Singapore. 


 }
}

\markboth{Journal of \LaTeX\ Class Files,~Vol.~14, No.~8, August~2021}%
{Shell \MakeLowercase{\textit{et al.}}: A Sample Article Using IEEEtran.cls for IEEE Journals}


\maketitle

\begin{abstract}

The increasing demand for communication is degrading the electromagnetic (EM) transmission environment due to severe EM interference, significantly reducing the efficiency of the radio frequency (RF) spectrum. Metasurfaces, a promising technology for controlling desired EM waves, have recently received significant attention from both academia and industry. However, the potential impact of out-of-band signals has been largely overlooked, leading to RF spectrum pollution and degradation of wireless transmissions. To address this issue, we propose a novel surface structure called the Filtering Reconfigurable Intelligent Computational Surface (FRICS). We introduce two types of FRICS structures: one that dynamically reflects resonance band signals through a tunable spatial filter while absorbing out-of-band signals using metamaterials and the other one that dynamically amplifies in-band signals using computational metamaterials while reflecting out-of-band signals. To evaluate the performance of FRICS, we implement it in device-to-device (D2D) communication and vehicular-to-everything (V2X) scenarios. The experiments demonstrate the superiority of FRICS in signal-to-interference-noise ratio (SINR) and energy efficiency (EE). Finally, we discuss the critical challenges faced and promising techniques for implementing FRICS in future wireless systems.
\end{abstract}


\section{Introduction}



\IEEEPARstart{M}{etasurface} is a two-dimensional planar structure composed of meta-atoms arranged in a specific pattern, typically with a thickness smaller than the wavelength of electromagnetic (EM) waves. This structure exhibits unique EM properties, offering flexible and efficient control over various characteristics of EM waves, such as polarization, amplitude and phase. 
In this context, reconfigurable intelligent metasurfaces (RISs) have recently emerged as a representative technology capable of managing wireless signal propagation and enhancing communication quality~\cite{marco1}. For example, in scenarios where users experience dead zones, implementing RISs, especially at the periphery of a cell, can create propagation paths that enable beamforming, thereby improving signal reception~\cite{editor1}. 


Although RISs are advancing as a technology for creating intelligent wireless RF environments, there remains a critical need to enhance their capability to adapt to complex radio frequency (RF) communication environments where multiple signals (e.g., the desired signals and interfering signals) overlap through passive reflection~\cite{TVT}. Addressing this challenge, Yang \textit{et al.} proposed an innovative approach integrating task-oriented computing and communication functions using computational metamaterials~\cite{yangReconfigurableIntelligentComputational2023}. They introduced a novel metasurface structure known as reconfigurable intelligent computational surface (RICS), designed to enable dynamic tunable signal reflection and amplification via computational metamaterials. However, in complex RF environments with significant interference, both RICS and traditional RISs struggle to effectively mitigate interferences. For instance, when signals out-of-band interact with a metasurface, they often undergo unpredictable and uncontrolled reflection. This unmanaged reflection sometimes will exacerbate interference issues, impacting receiver efficiency and causing severe spectrum pollution, ultimately degrading overall communication system performance.


Various methods have been explored to achieve interference cancellation in wireless communications. In passive RISs, such as those discussed by Tang \textit{et al}.~\cite{tangWirelessCommunicationsReconfigurable2021}, interference elimination is achieved by utilizing signal reflection from the RISs. However, passive RISs can only degrade interference from the front side of the metasurface and are ineffective against interference from the back side.  To address this limitation, STAR-RIS proposed by the authors in~\cite{STAR-RIS} divides the incident signal into two components, allowing for both reflection and refraction. This capability enables interference elimination from both the front and back sides through phase shift optimization. Nonetheless, these methods often result in significant signal attenuation after passing through the metasurface, thereby reducing the effectiveness of interference cancellation. In contrast, active RISs, as discussed in~\cite{activeRIS}, integrate amplifier components directly onto the metasurface. This design enhances the amplitude of the incident signal, thereby increasing signal power and maximizing interference cancellation at the receiving end. However, the incorporation of amplifiers increases energy consumption, which can impact the overall energy efficiency of the system. 


To tackle the challenges previously discussed, we propose a novel filtering reconfigurable intelligent computational surface, called FRICS, which can leverage absorbing metamaterials to effectively absorb interference and computational metamaterials to amplify the desired signal. The four types of metasurfaces are compared in Table \ref{tab:t1}. We observe that passive RISs leverage reflection to cancel interference and STAR-RIS enhances cancellation by combining reflection and refraction, active RISs offer better interference cancellation through signal amplification despite potential drawbacks in energy efficiency. In contrast with them, the innovative configuration enables FRICS to achieve comprehensive interference elimination while simultaneously reducing energy consumption. Compared to existing interference cancellation methods, FRICS enhances both the signal-to-interference-plus-noise ratio (SINR) and energy efficiency (EE) of the system.

\begin{table*}[]
\renewcommand\arraystretch{1.25} 
\captionsetup{font={small}}
\caption{Comparison of different metasurfaces designs}
\label{tab:t1}   
\resizebox{\textwidth}{!}{  
\begin{tabular}{ccccccc}    
\toprule
 &filtering & frequency adjustment & Simultaneous reflection and transmission  & signal amplification  & power consumption & interference cancellation capability\\
\midrule    
Passive RIS\cite{tangWirelessCommunicationsReconfigurable2021} 
&\usym{2715} &\usym{2715} & \usym{2715} & \usym{2715} & middle &  low \\
STAR-RIS\cite{STAR-RIS} & \usym{2715} & \usym{2715} & \usym{1F5F8} & \usym{2715} & middle& low\\
Active RIS \cite{activeRIS} & \usym{2715} & \usym{2715} & \usym{2715} & \usym{1F5F8} & high & middle \\
proposed FRICS & \usym{1F5F8} & \usym{1F5F8} & \usym{1F5F8} & \usym{1F5F8} & low & high\\
\bottomrule   
\end{tabular}
}
\end{table*}

\begin{figure*}
    \centering
     \captionsetup{font={small}}
    \includegraphics[width=0.95\linewidth]{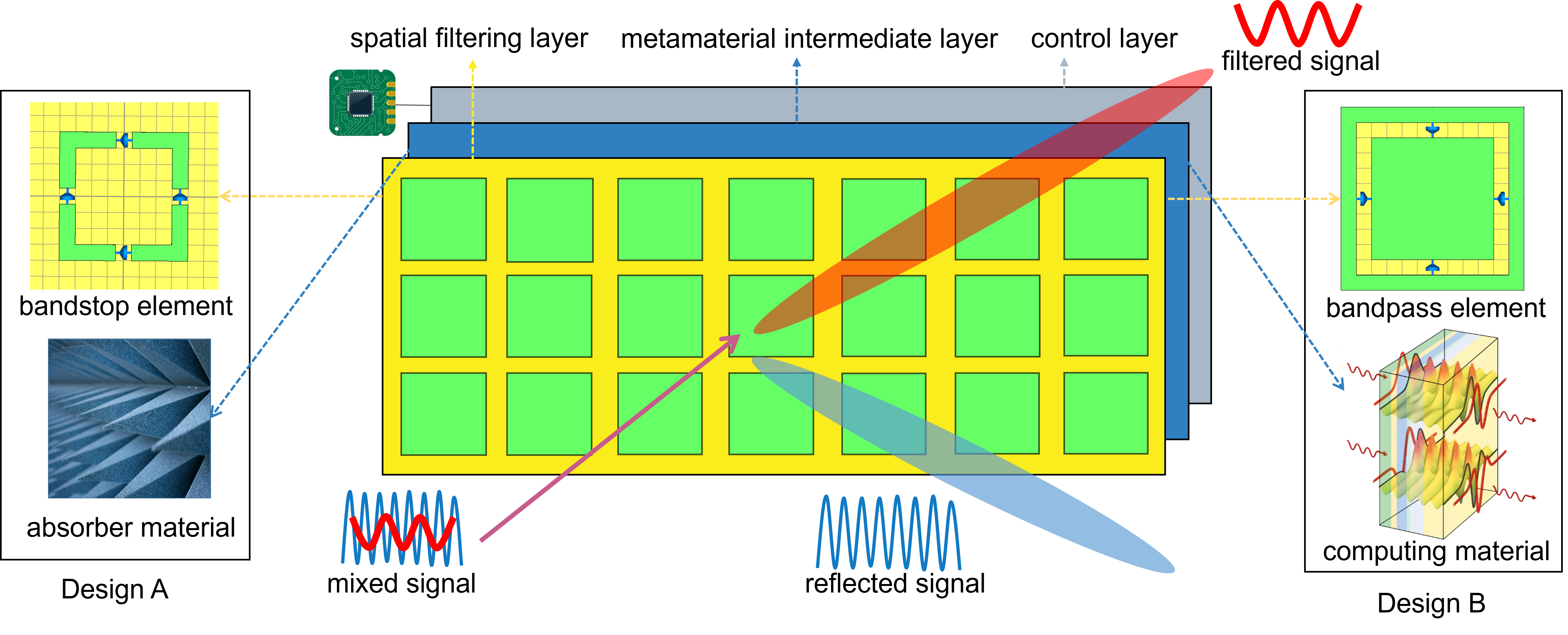}
    \caption{The structure of FRICS. It consists of a spatial filtering layer, an intermediate layer made of metamaterials, and a control layer. Design A on the left includes a bandstop filter layer plus an absorbing material layer. Design B on the right includes a bandpass filter layer plus a computational metamaterial layer.}
    \label{fig:FRICS}
\end{figure*}

\section{Fundamentals of FRICS}
As shown in Fig.~\ref{fig:FRICS}, our proposed FRICS comprises three layers: a spatial filtering layer, a metamaterial intermediate layer, and a control layer. The control layer includes a circuit board hosting an intelligent controller responsible for adjusting the tunable parameters of the spatial filtering layer. This adjustment is commonly facilitated by a field-programmable gate array (FPGA).
In the subsequent sections, we will detail the structure of the spatial filtering layer and explore its collaborative design with the metamaterial intermediate layer to achieve RF spectrum purification in typical application scenarios.

\subsection{Spatial Filtering Layer}
Based on the generalized Huygens' principle and the theory of metamaterial equivalent circuit models \cite{huygens}, a metasurface can be conceptually represented as a series connection of capacitive grids and inductive grids, thereby forming a series LC resonant circuit. This equivalence arises because the metal resonators within the metasurface structure can be analogously considered inductors. In contrast, the gaps or spaces between these resonators can be analogously treated as capacitors in an electrical circuit model.

\begin{figure*}
	\centering
 \captionsetup{font={small}}
	\begin{subfigure}{1.0\linewidth}
		\includegraphics[width=\linewidth]{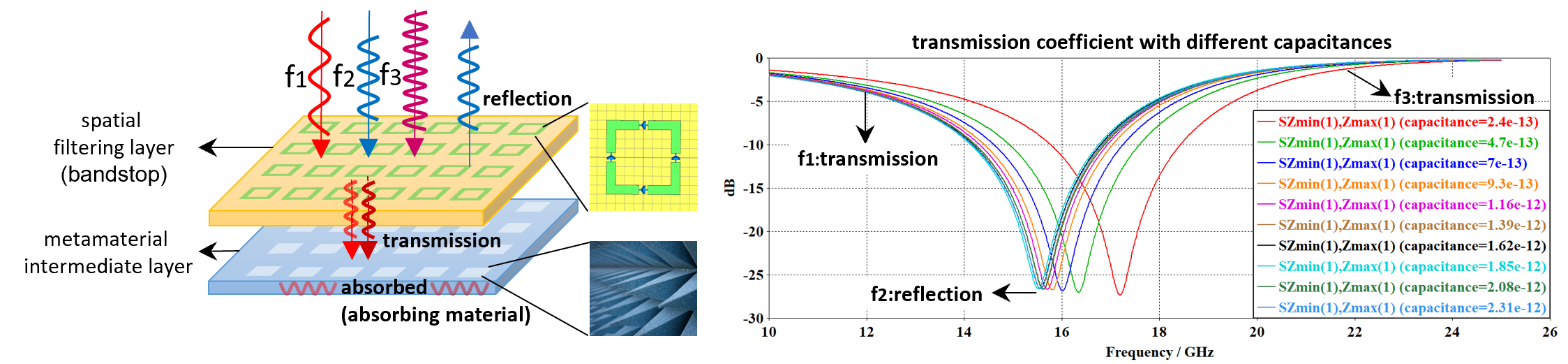}
		 \caption{}
		\label{fig:subfigA}
	\end{subfigure}
	\begin{subfigure}{1.0\linewidth}
		\includegraphics[width=\linewidth]{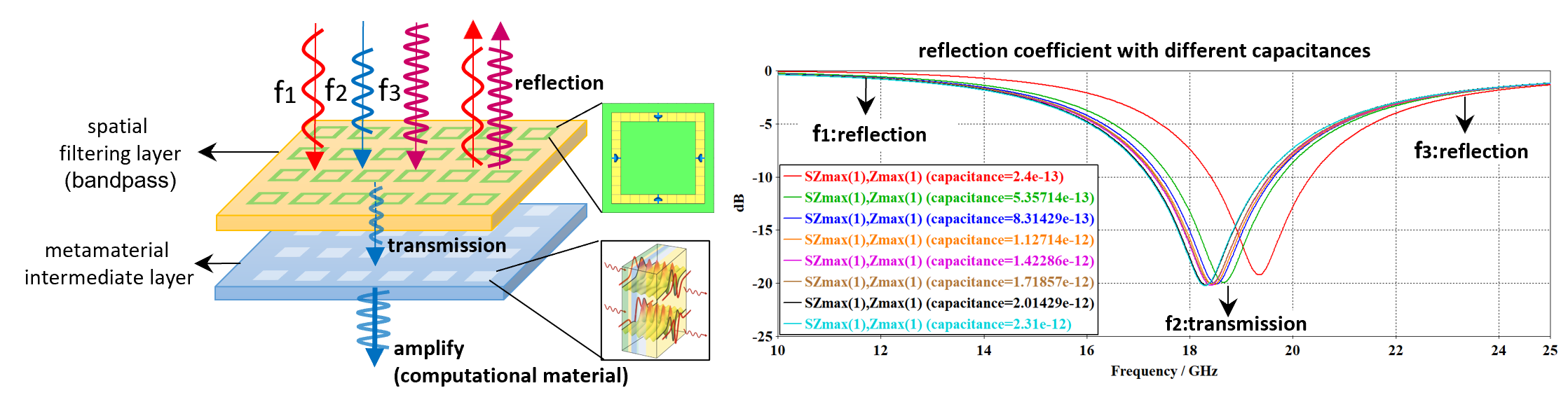}
		 \caption{}
		\label{fig:subfigB}
	\end{subfigure}
	 \caption{The left shows the signal propagation of the FRICS, and the right shows the CST simulation results of the spatial filtering layer, where the design A of FRICS is in (a), and the design B of FRICS is in (b).}
	\label{fig2}
\end{figure*}

In the equivalent LC resonant circuit, the capacitive reactance of the equivalent capacitance decreases with increasing frequency, while the inductive reactance of the equivalent inductance increases. This behavior impacts the performance of ring-shaped elements in the circuit, which exhibit a relatively high shunt impedance, thus behaving similarly to an open circuit. Consequently, EM waves can effectively pass through the FRICS at both higher and lower frequencies.
However, the transmission performance of the FRICS is weakest in the intermediate frequency range, resulting in a bandstop characteristic, as illustrated in Fig.~\ref{fig2}(a). At intermediate frequencies, the combination of the decreasing capacitive reactance and the increasing inductive reactance creates conditions that inhibit the efficient transmission of EM waves, leading to a notable dip in performance. 


On the other hand, as illustrated in Fig.~\ref{fig2}(b), aperture-shaped elements exhibit the opposite behavior to ring-shaped elements, resulting in a bandpass characteristic. This means that EM waves can pass through the structure more effectively at certain intermediate frequencies, rather than being blocked. The unit structure of the spatial filtering layer of Design B is depicted on the left side of Fig.~\ref{fig2}(b). In this figure, the yellow color represents the substrate, the green color represents the metal patches, and the blue color represents the variable capacitor diodes. These diodes can adjust their capacitance values, which in turn influences the transmission characteristics of the spatial filtering layer. The right figure in Fig.~\ref{fig2}(b) shows the transmission coefficient of the spatial filtering layer with different capacitance values of the variable capacitor diodes. By varying the capacitance, the transmission properties of the filter can be tuned, allowing for the selective passage of certain frequency ranges while blocking others. This tunability is key to achieving the desired bandpass characteristic, enabling the filter to efficiently manage the passage of EM waves based on their frequency.

In practical scenarios, the frequencies of the desired signal and the interfering signal can vary due to the communication environment. To address this, we employ variable capacitors, commonly referred to as tuning diodes, to dynamically adjust the operating frequency range of the FRICS by controlling the bias voltage. This allows the metasurface to be integrated into the spatial filtering layer to achieve frequency tuning.
Specifically, when a reverse bias voltage is applied to the variable capacitor, its capacitance changes accordingly. In this paper, we utilize the variable capacitor SMV2019-079LF as an example~\cite{8445639}, in which the capacitance changes with the reverse bias voltage. For example, when the bias voltage of the diode changes from 0V to -19V, the capacitance of the diode changes from 2.31pF to 0.24pF. Consequently, the tunable control layer can dynamically adjust the reflection frequency range of the FRICS by controlling the reverse bias voltage of each unit using a FPGA. We further explore the CST simulations and we vary the values of the RLC series circuit to simulate the electrical characteristics of the variable capacitor under different bias voltages. By implementing this dynamic tuning capability, the FRICS can effectively adapt to varying signal frequencies, enhancing its performance in diverse communication environments. 
\vspace{-2mm}
\subsection{Metamaterial Intermediate Layer}
To fully leverage the potential of the metamaterial intermediate layer, it is crucial to appropriately design the surface structure. This can be achieved by using \textbf{absorbing materials to absorb interference signals} or employing \textbf{computational materials to amplify desired signals}.  Specifically, as a novel type of material that can significantly weaken or absorb EM wave energy, absorber metamaterials can help to reduce EM interference~\cite{10.1063/1.3276072}. Additionally, absorber metamaterials possess characteristics such as being lightweight, corrosion-resistant, and temperature-resistant, making them suitable for deployment in wireless communication environments. According to the material compositions, the common absorber materials can be classified into carbon-based absorber materials, iron-based absorber materials, ceramic-based absorber materials, and chiral materials. In our design, iron oxide material is implemented as the intermediate layer due to its high magnetic permeability, high resistivity, and good impedance matching characteristics. For example, $\rm \ce{BaCo_{0.9}Si_{0.95}Fe_{10.1}O_{19}}$ exhibits an absorption rate of up to 99\% within the frequency range of 11-13 GHz~\cite{ABBAS200720}. On the other hand, computational metamaterials refer to a specially designed block of metasurface that allows mathematical operations on the incident waves, such as spatial differentiation, integration, and convolution~\cite{yangReconfigurableIntelligentComputational2023}. To be specific, a computational metamaterial can achieve amplification of the incident signal that impings the FRICS. Therefore, leveraging the properties of the computational metamaterial can enhance the signal-to-noise ratio by amplifying the desired signal while avoiding additional energy consumption.

To sum up, by incorporating absorbing materials, the intermediate layer can effectively mitigate unwanted/interfering signals, thereby enhancing the overall signal quality. Meanwhile, using computational materials can help to amplify the desired signals, improving their strength and clarity. This dual capability of absorption and amplification makes the metamaterial intermediate layer a vital component in the FRICS design, enabling it to adapt to various signal environments and maintain efficient communication. Therefore, the intermediate layer's design should be tailored to the specific requirements of the application, ensuring optimal performance of the FRICS system.



\subsection{FRICS Architecture Design}
According to the configuration of the spatial filtering layer and the metamaterial intermediate layer, FRICS has two designs, which are described as follows.

\begin{itemize}
    \item \textbf{FRICS Design A: bandstop filtering layer + absorbing intermediate layer}. In this design, the spatial filtering layer achieves a bandstop filtering effect while an absorber material is used as the metamaterial intermediate layer.
     As simulated using CST in the right of Fig.~\ref{fig2}(a), the transmission coefficient exhibits a notable decrease within the frequency range of 16 GHz to 18 GHz, indicating reflective characteristics. In the remaining frequency range, the transmission coefficient remains relatively unchanged, demonstrating transmission characteristics. Therefore, as shown in the left figure of Fig.~\ref{fig2}(a), the in-band signal ($f_{2}$) can be reflected and the out-of-band signals ($f_{1}$ and $f_{3}$) can be passed through the filtering layer. Then absorbing material can absorb these signals to achieve interference cancellation.
\end{itemize}

 \begin{itemize}
     \item \textbf{FRICS Design B: bandpass filtering layer + computational intermediate layer}. In this design, the spatial filtering layer achieves a bandpass filtering effect while the computational metamaterial is used as the intermediate layer. As illustrated in Fig.~\ref{fig2}(b), there is a notable decrease in the reflection coefficient within the frequency range of 17 GHz to 19 GHz, indicating transmission characteristics. In the out-of-band range, the reflection coefficient remains relatively unchanged, demonstrating reflective characteristics. Consequently, as illustrated in the left figure of Fig.~\ref{fig2}(b), the desired signal ($f_{2}$) passes through the spatial filtering layer and then can be amplified with the aid of computational metamaterials. Meanwhile,  the out-of-band signals ($f_{1}$ and $f_{3}$) can be reflected through the filtering layer.
 \end{itemize}

Based on the two designs of FRICS, for the communication scenario where the desired signal and interfering signal coexist, the proposed FRICS can effectively separate the desired signals from the interfering signals. Meanwhile, the unwanted interfering signals can be absorbed and the desired signal can be amplified by configuring the design of FRICS appropriately, thereby improving energy efficiency.

\section{Applications and Performance Evaluation}
Based on the two designs introduced above, we validate the superiority of FRICS in two typical scenarios. 
\subsection{Application Scenarios}

\begin{figure*}
	\centering
    \captionsetup{font={small}}
    \begin{subfigure}{0.47\textwidth}
    \centering
        \includegraphics[width=\textwidth]{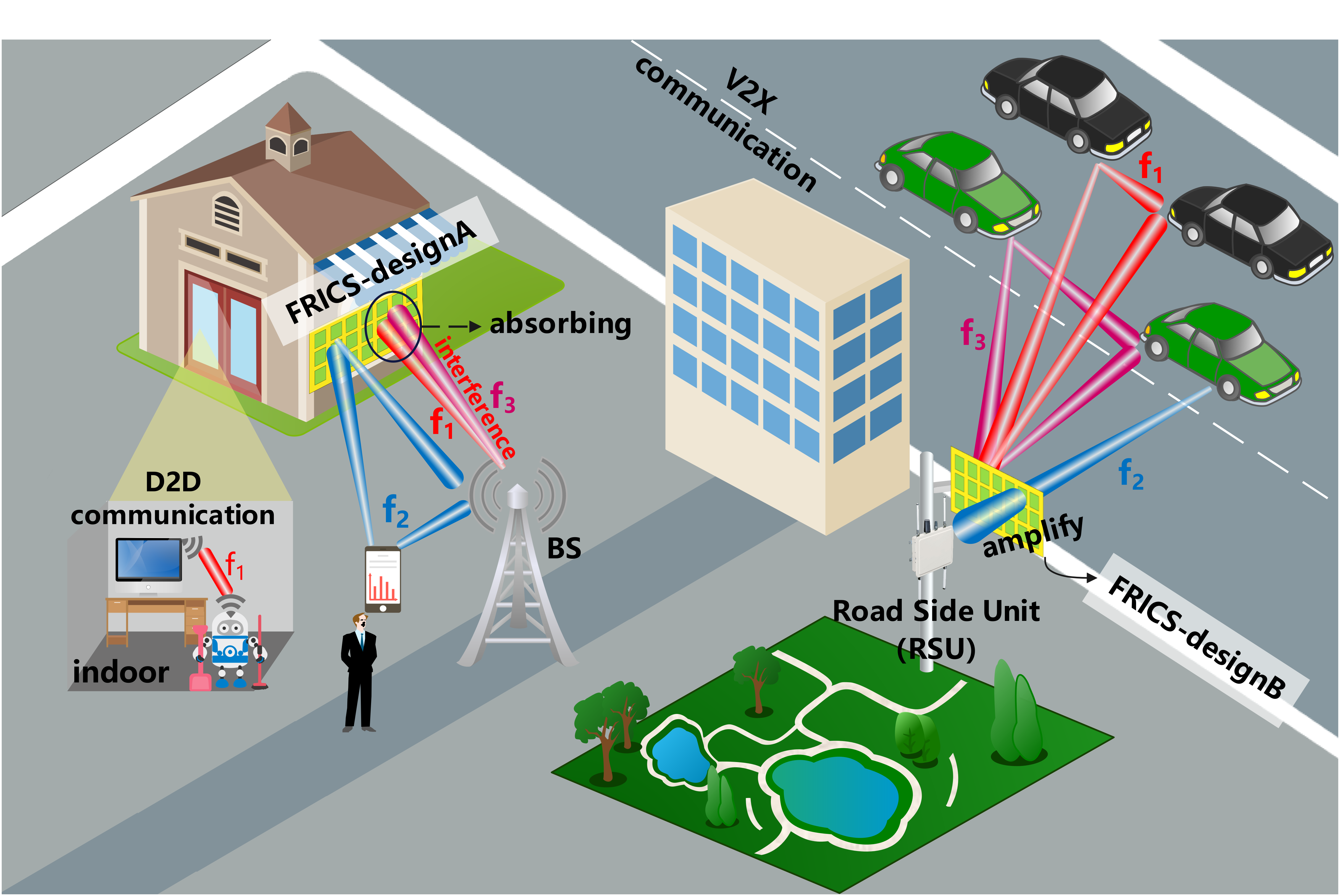}
        \caption{}
        \label{fig:designFRICS}
    \end{subfigure}
    \hfill
    \begin{subfigure}{0.47\textwidth}
    \centering
        \includegraphics[width=\textwidth]{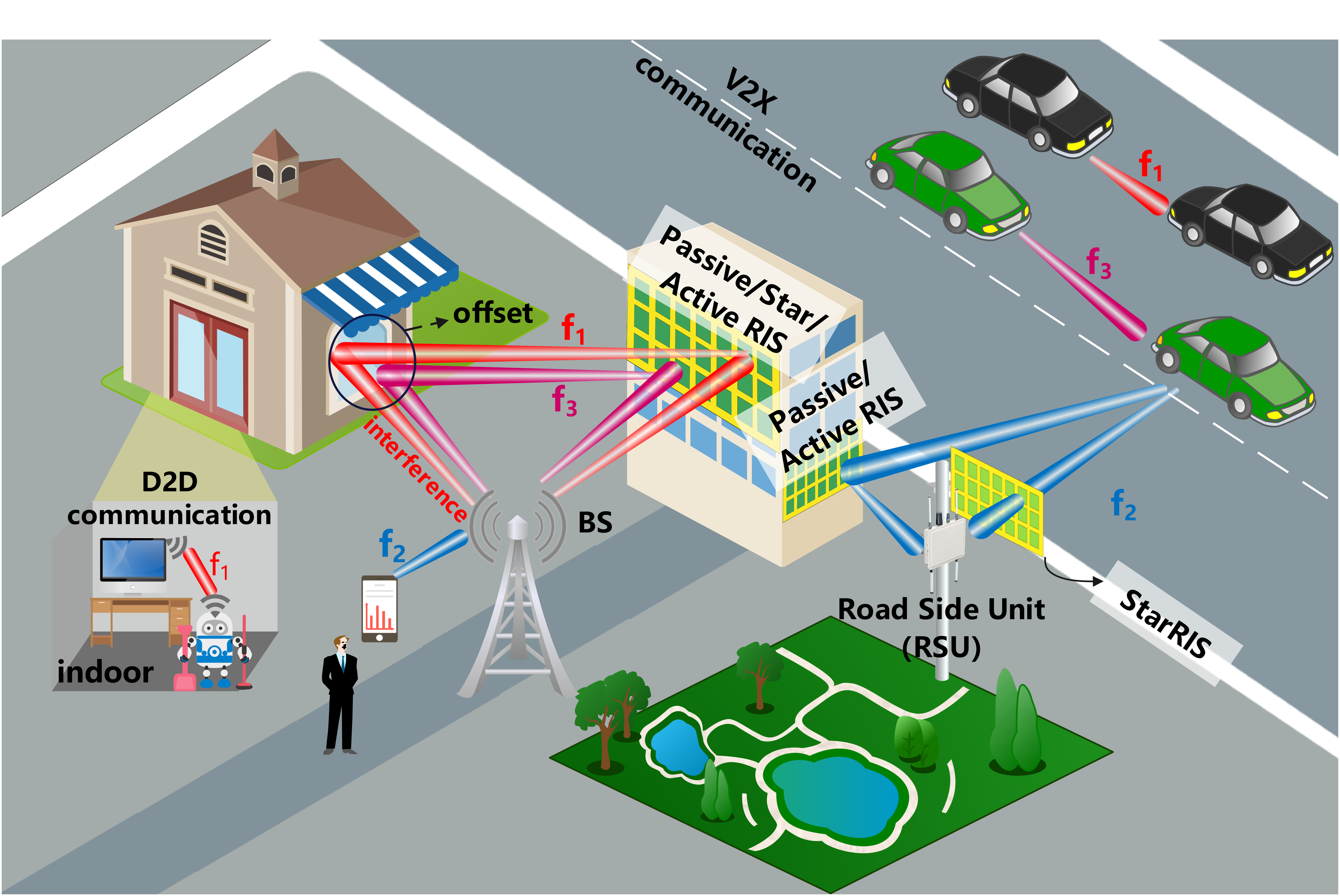}
        \caption{}
        \label{fig:designRIS}
    \end{subfigure}
	\caption{In (a), the application scenario of FRICS with Design A is shown on the left, wherein the outdoor interference signal at the frequency $f_1$ or $f_3$ is filtered and absorbed via FRICS. Concurrently, the signal $f_2$ is reflected to enhance wireless link quality. In contrast, passive/active/star-RIS can only compensate for interference signals by adjusting the reflection phase, as shown on the left of (b). The typical application scenario of FRICS with Design B is depicted on the right of (a), where the desired signal from the vehicle to the roadside unit (RSU) via $f_2$ can be filtered and amplified by FRICS. Meanwhile, the signals $f_1$ and $f_3$ are reflected to enhance the V2V transmission. For comparison, passive/active/star-RIS just reflects/refracted the desired signals to improve the receiving quality, as shown in the right of (b).}
	\label{fig:design}
\end{figure*}

\subsubsection{\textbf{Interference signal filtering and absorbing}}
As illustrated in Fig. \ref{fig:design}(a), we examine a device-to-device (D2D) communication scenario for efficient transmission, where a D2D indoor communication pair operating at the frequency $f_1$ and may experience significant co-frequency interference from outdoor cellular communications. To eliminate the co-frequency interference from outdoor cellular communication, the FRICS is implemented on the exterior wall of the building. This setup allows the incoming interfering signal at $f_1$ or $f_3$ to pass through the spatial filtering layer of the FRICS. Once through this layer, the signal is absorbed by the metamaterial intermediate layer, which consists of absorbing materials designed to mitigate interference. At the same time, the signal between the users and the base station (BS) at $f_2$ will be reflected to enhance the communication quality. It is observed that the implementation of FRICS plays a crucial role in maintaining high-quality indoor communication by mitigating external interferences.

\subsubsection{\textbf{Desired signal filtering and amplification}} As illustrated in Fig. \ref{fig:design}(b), we consider a vehicular-to-everything (V2X) scenario, where a vehicle is communicating with a roadside unit (RSU) at frequency $f_2$. Concurrently, a vehicle-to-vehicle (V2V) pair transmits at frequencies $f_1$ or $f_3$. To improve the SINR for the RSU, when the uplink signal from the vehicle impinges the FRICS, the signal is allowed to go through the filtering layer and then amplified through computational metamaterial. This amplification ensures that the uplink signal is strong and clear when it reaches the RSU, thereby boosting the intended signal's strength and thus enhancing communication reliability and efficiency. At the same time, V2V communication is occurring. The FRICS can reflect the V2V communication signals, thereby enhancing their communication quality. By reflecting these signals, the FRICS helps to maintain robust and uninterrupted V2V communication, which is crucial for safety and coordination between vehicles.

\begin{figure*}[htbp]
    \centering
    \captionsetup{font={small}}
    \begin{subfigure}{0.48\textwidth}
        \centering
        \includegraphics[width=\textwidth]{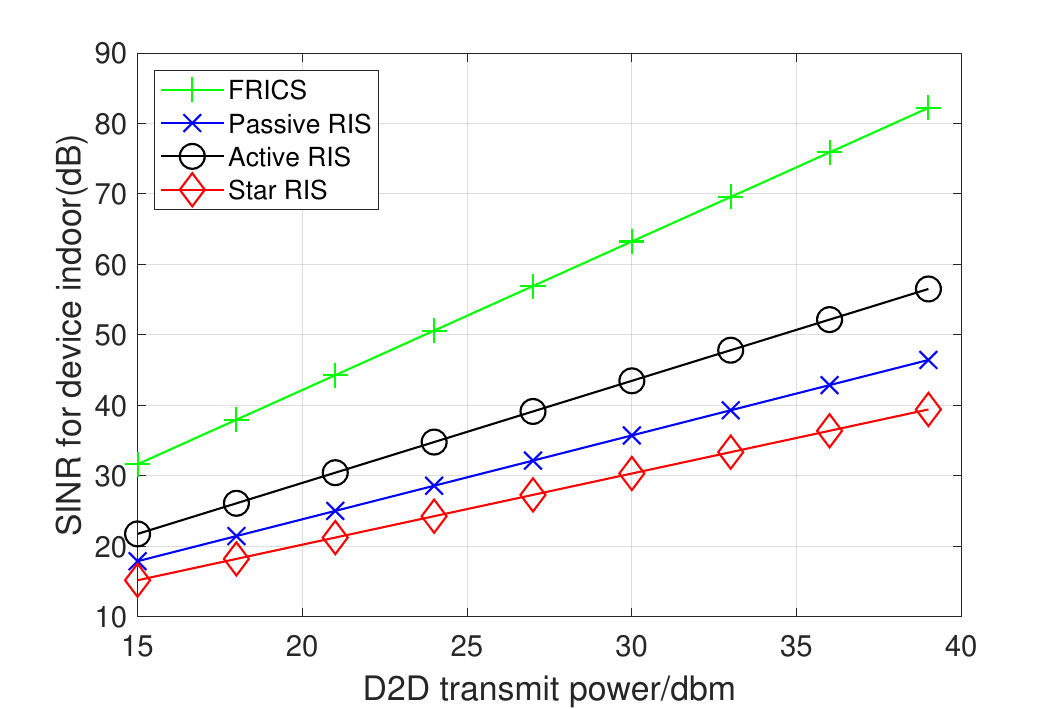}
        \caption{}
        \label{fig:sinrdesignA}
    \end{subfigure}
    \hfill
    \begin{subfigure}{0.48\textwidth}
        \centering
        \includegraphics[width=\textwidth]{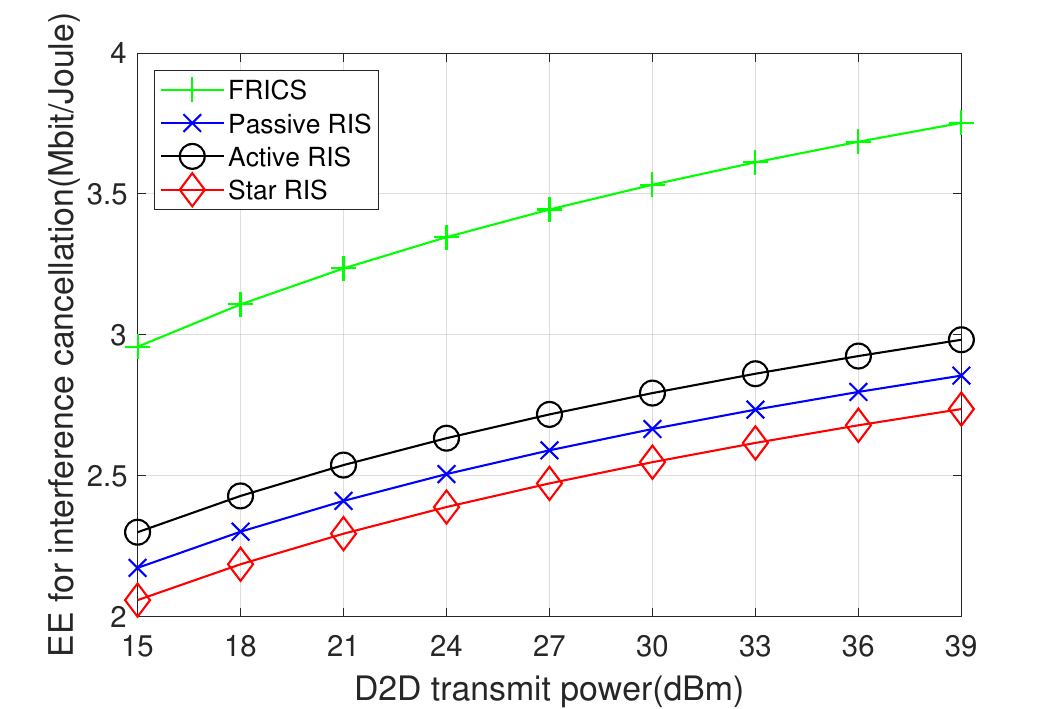}
        \caption{}
        \label{fig:eedesignA}
    \end{subfigure}
    \hfill
    \begin{subfigure}{0.48\textwidth}
        \centering
        \includegraphics[width=\textwidth]{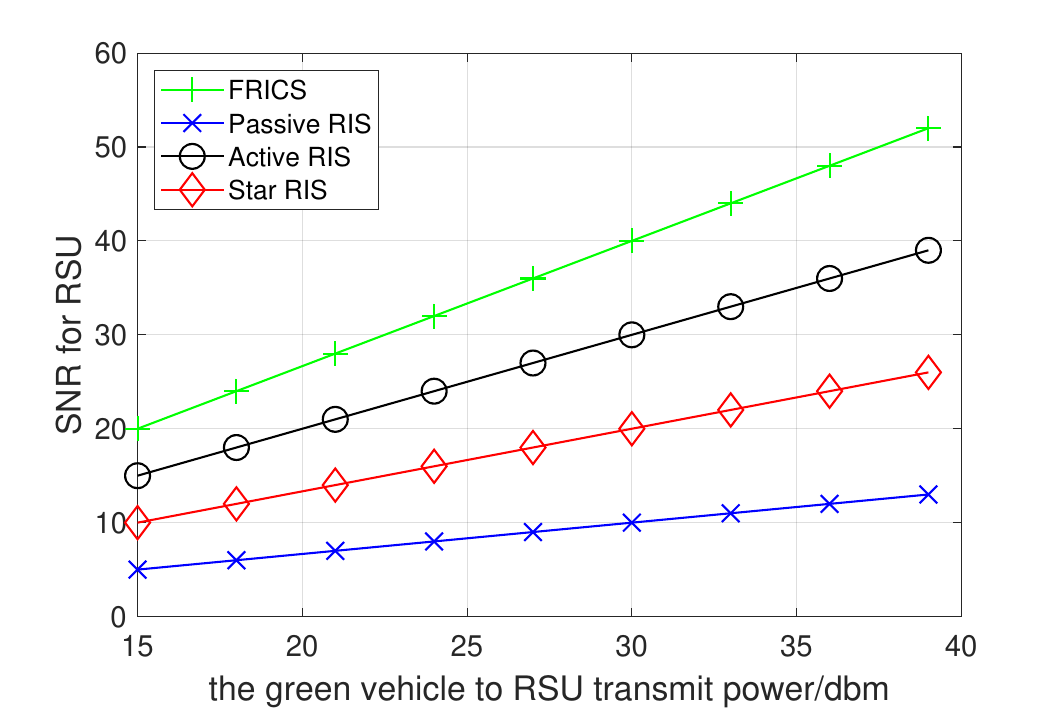}
        \caption{}        
        \label{fig:sinrdesignB}
    \end{subfigure}
    \hfill
    \begin{subfigure}{0.48\textwidth}
        \centering
        \includegraphics[width=\textwidth]{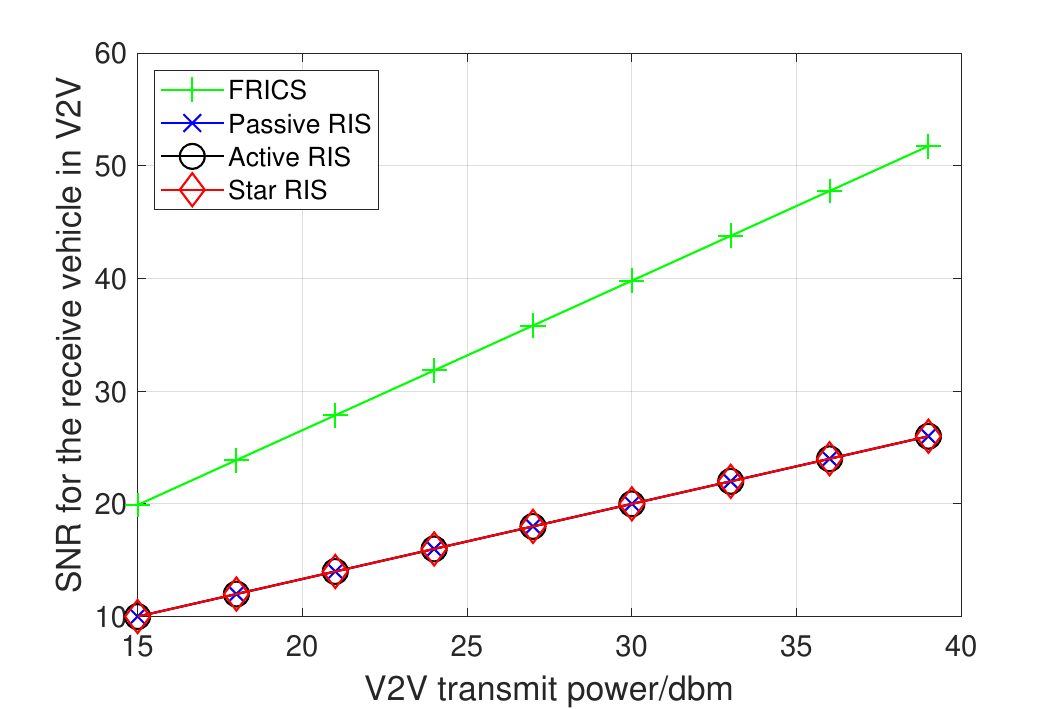}
        \caption{}
        \label{fig:sinr2designB}
    \end{subfigure}
    \caption{The comparison of SINR and EE for the receiving device in an indoor environment after passing through FRICS (operating with Design A), passive RIS, active RIS, and star RIS (with simultaneous transmission and reflection mode) is shown in (a) and (b), respectively. The comparison of SNR for the RSU and the V2V communication after passing through FRICS (operating with Design B), passive RIS, active RIS, and star RIS (with full transmission mode) is presented in (c) and (d), respectively. }
    \label{fig:experiment}
\end{figure*}

\subsection{Simulation Results and Discussions}
The configuration parameters for the spatial filtering layer are shown in Table \ref{tab:parameters}. To demonstrate the benefits of FRICS, we evaluated energy efficiency with the bandwidth set to $1$ MHz. 
To accurately compare the energy consumption of different metasurfaces designs with interference mitigation when using variable capacitors, the power consumption of the RIS reflection units is $0$, the power consumption of the driver generator is $250$ mW, the power consumption of the bias voltage amplifier is $180$ mW, and the FPGA power consumption is $1.5$ W~\cite{wangReconfigurableIntelligentSurface2024}.

\subsubsection{Simulation results of Design A}

\begin{table}[thb]
\captionsetup{font={small}}
\caption{Experimental Parameters of FRICS}
\label{tab:parameters}
\begin{tabular}{|m{0.14\textwidth}|m{0.08\textwidth}||m{0.1\textwidth}|m{0.06\textwidth}|}
\hline
\textbf{Parameters}& \textbf{Values}  &  \textbf{Parameters}   & \textbf{Values} \\
\hline
Substrate side length  & 6mm & Patch width & 1mm\\

Substrate thickness & 0.38mm&Pore size& 1mm \\

Substrate material&FR-4& Patch material& Copper\\

diode capacitance &0.24-2.31pF & Patch thickness&0.038mm\\
\hline
\end{tabular}
\end{table}

In this scenario, we compare the SINR and EE of our proposed FRICS with the latest RIS models (passive RIS, and active RIS and star RIS) in terms of interference mitigation. The D2D transmission power ranges from 15 dBm to 39 dBm. For the communication simulation scenarios, we assume a D2D communication distance of 10 meters and a cellular communication distance of 20 meters.


As illustrated in Fig. \ref{fig:experiment}(a), a comparison is presented between the SINR after interference mitigation utilizing FRICS, passive RIS, active RIS, and star RIS at the D2D receiver indoor. It is observed that FRICS achieves the highest SINR due to its ability to completely absorb the interfering signals. This superior performance can be attributed to the efficient absorption capabilities of the metamaterial intermediate layer, which effectively eliminates interference. In comparison, passive RIS experiences a greater degree of signal attenuation when eliminating interference, which ultimately affects its ability to eliminate interference signals. The star RIS, which can simultaneously transmit and reflect signals, even experiences a significant reduction in reflected signals compared to passive RIS, which affects its interference mitigation performance. Active RIS, on the other hand, amplifies the signals by incorporating additional amplifying circuits, resulting in stronger signals reaching the receiving end and effectively mitigating interference, thus achieving a higher SINR. 

Fig. \ref{fig:experiment}(b) illustrates the energy efficiency of FRICS for interference cancellation. As the D2D transmit power increases, the SINR at the D2D communication receiver can be improved, resulting in a gradual increase in the energy efficiency for interference cancellation. Notably, the energy efficiency of FRICS remains the highest due to its ability to effectively cancel interference without introducing additional energy, resulting in near-zero power consumption for the driving generator. Given the related power consumption parameters, the energy efficiency of amplifying signals for active RIS (with the amplification factor of 2 and each reflection unit consumes 150 mW of energy) is higher than the others.

\subsubsection{Simulation results of Design B}
In this scenario, we evaluate the SNR achieved at the RSU when the transmission power of the vehicle ranges from 15 dBm to 39 dBm. We assume that the gain of star RIS and FRICS is 2.

Fig. \ref{fig:experiment}(c) illustrates a comparison between the FRICS and the benchmark RISs designs in terms of SNR for V2X communication, with the transmission power of the vehicle ranging from 15 to 39 dBm. We observe that for the RSU, the FRICS achieves the highest SNR performance. This is because the computational metamaterials amplify the desired signal. In contrast, passive RIS has no signal amplification capability, resulting in the worst performance. Similarly, Fig. \ref{fig:experiment}(d) illustrates that only FRICS can extend the performance of V2X by amplifying in-band signal while improving the quality of V2V communication via out-of-band signal reflection.

To sum up, we observe from Fig. \ref{fig:experiment}(a)-(d) that the proposed FRICS demonstrates superior interference mitigation capabilities without the need for additional energy input, achieving the highest SINR/SNR among the compared benchmark schemes. Its ability to absorb interfering signals and amplify the desired signals completely provides a significant advantage in maintaining high communication quality and energy efficiency.

\section{Challenges and Future Directions}
The introduction of the FRICS designs has significant implications for future wireless communication systems, particularly in the context of upcoming sixth-generation (6G) networks with a more complicated RF communication environment.
In the following, we outline the challenges and potential directions, which aim at paving the way for more feasible solutions to achieving RF spectrum purification in future 6G networks.

\subsection{FRICS Design for Energy Harvesting} In this paper, we have utilized absorbing materials as the intermediate layer to achieve interference cancellation. However, we believe that the metamaterial intermediate layer holds even greater potential. By implementing energy-conversion metamaterials, such as photovoltaic conversion metamaterials, as the intermediate layer, we can achieve energy harvesting when the filtered signals, including interference signals, impinge on the intermediate layer~\cite{powertransfer}.  
This approach enhances the deployment flexibility of the metasurfaces by enabling the collection of energy from various sources, such as solar power or wireless signals. It eliminates the need for additional power sources, allowing metasurface deployment on various object surfaces. This capability can significantly extend the sustainability of the metasurfaces, making them more adaptable to diverse environments and applications.

Looking ahead, there are opportunities to explore different types of metasurfaces to achieve additional functionalities. For example, time-targeting metasurfaces can dynamically adjust their properties based on temporal changes, enabling precise control over when certain signals are absorbed or amplified.
By integrating such advanced functionalities, future metasurfaces can offer even greater capabilities and versatility, paving the way for innovative applications in wireless communication, energy harvesting, and beyond.

\subsection{FRICS Design for Phase Modulation} We have observed that although FRICS can effectively eliminate interference signals, controlling the phase of the reflected signals remains a challenge since precise phase control is crucial for several metasurfaces-based applications such as beamforming. 
Recent works have achieved filtering while adjusting the phase~\cite{liangFilteringReconfigurableIntelligent2024}. However, their multi-layer structures and additional amplifiers consume significant energy, which violates the principle of metasurface design. Moreover, they ignore the issue of same-frequency interference, which is more serious in practice. Therefore, it holds promise to investigate how to directly modulate the phase of FRICS. 

Furthermore, when we consider the coexistence of the FRICS and existing RISs within the same physical space, the functionalities provided by FRICS can complement the limitations of traditional RISs. For instance, when the metamaterial intermediate layer of FRICS employs optoelectronic conversion materials, the energy obtained by the FRICS can not only sustain its operation but also serve as an energy supplement for other RISs. Meanwhile, the beamforming capability of the RISs can compensate for the limitations of FRICS by enabling phase adjustment.

\subsection{Interplay between FRICS Design and ISAC} Integrated sensing and communications (ISAC) has emerged as a crucial technology in the context of 6G aiming to achieve both perception and communication~\cite{marco2}. Traditional approaches to integrated sensing often prioritize either communication-centric or perception-centric designs, making it challenging to achieve a unified ISAC system. However, the proposed FRICS holds the potential to address this challenge by filtering signals in different frequency bands, allowing the reflected signals to be used for communication and the filtered signals for sensing. For instance, the signal filtered to the intermediate layer of FRICS can directly serve as information for environmental sensing, such as humidity, pollutant density, and more. 

Furthermore, to effectively suppress interferences using the metamaterial intermediate layer of FRICS, the transmission coefficient of FRICS should be configured appropriately based on the frequency of signals. To obtain this frequency information, ISAC may play a crucial role in detecting the frequency of the signals, e.g., with the aid of deep learning methods~\cite{TVT}. This dual functionality enables FRICS to enhance the capabilities of ISAC systems, providing a seamless integration of communication and sensing tasks within a single framework. 

\subsection{Low-Energy and Low-Complexity Transceiver Design} 
In traditional wireless communication systems, the receiver's antenna captures wireless signals, which then require filtering to extract the desired signal. However, these conventional designs often result in increased receiver complexity and higher energy consumption.
In contrast, the proposed FRICS offers several advantageous characteristics, including low energy consumption and tunability. By integrating FRICS into receiver designs, it mitigates receiver complexity and reduces energy consumption significantly.
Moreover, FRICS facilitates dynamic adjustment of the filtering frequency, which aligns with principles of green and low-carbon design. This capability not only provides enhanced flexibility but also holds the potential to enhance communication performance.

If we go one step further and suppose that the receiver is surrounded by multiple FRICS unis, by assigning different operating frequencies to each directional FRICS, we can selectively ensure that the desired signal is received while filtering out interfering signals. This approach leverages the directional capabilities of FRICS to achieve significant improvements in signal reception quality, interference management, and overall communication performance.

\section{Conclusion}
This paper investigated a novel filtering RICS (FRICS) structure and introduced two different designs. Design A uses a tunable spatial filter to dynamically reflect signals within the resonance band while absorbing out-of-band signals. Design B enables the transmission and amplification of the desired signals within the resonant band while reflecting out-of-band signals. Furthermore, this paper explores two typical application scenarios where FRICS can be employed and evaluates its performance in these contexts. Finally, we outline several challenges and opportunities, highlighting promising directions for future research in this field.


\bibliographystyle{IEEEtran}

\bibliography{IEEEabrv,reference} 

\end{CJK}
\end{document}